\documentstyle[12pt,psfig,aaspp4]{article}
\begin{document}



\title{QSO HOSTS AND COMPANIONS AT HIGHER REDSHIFTS}


\author{J. B. Hutchings}

\affil{Herzberg Institute of Astrophysics, NRC of Canada\\
Victoria, B.C., Canada}



\begin{abstract}

   This review presents the current state of work on QSO hosts and
companions at redshifts above 1. This includes the properties of
QSO host galaxies, such as size, scale length, and luminosity,
and morphology, as they appear to change with redshift and
radio activity. This leads to a view of how the properties of
galaxies that host QSOs change with cosmic time.
I also review studies of the galaxy companions to QSOs at higher redshifts,
and studies of the emission line gas in and around higher redshift QSOs.
These topics should see great progress in the next decade.

\end{abstract}



\section{Host galaxy detection}

     In the past decade we have made significant progress in
detecting and measuring the host galaxies of QSOs at redshifts 
in the range 1 to 2.5, and even above 3 in a few instances. This is 
partly due to the fact that the hosts at higher redshifts are in very active
star-forming stages of their lives, and are thus bright in the observed
rest-frame UV. The first significant paper on high redshift host galaxies
came from KPNO data with resolution only about 1.3 arcsec, showing dramatically
that radio-loud hosts were both bright and large (Heckman et al 1991).
Since then the availability of NIR imaging and higher resolution imaging 
(via HST or adaptive optics) have been responsible for the bulk of the
higher z investigations to date, and which I attempt to review here
(see e.g. Lowenthal et al 1995, Lehnert et al 1992, Hutchings 1995b, Aretxaga
et al 1998, Hutchings 1998, Hutchings et al 1999, Lehnert et al 1999,
Rix et al 2000, Kukula et al 2000, Ridgway et al 2001). 
It is clear that the use of 8-10m class telescopes will help us reach
to fainter details and higher redshifts where redshift dimming 
dominates, and will be a key to future investigations in this field
(Falomo et al 2000).

     There are thus a number of investigations that have reported on high
redshift host galaxies. Their interest as samples of high redshift
galaxies in general has somewhat been overshadowed by the recent
major progress in recognising and studying high redshift non-QSO galaxies,
via photometric redshifts and 10m telescope spectroscopy. However,
the QSO environment is still of considerable interest in the context of 
the general galaxy population at higher redshifts.

    As well as the redshifts above 1, I have selected host galaxy measurements
over the redshift range below 1 to cover the redshift range down to the
well-studied low redshift QSOs (Kotilainen et al 1998, Kotilainen
and Falomo 2000) . The diagrams show the result, attempting to be  
comprehensive at higher redshifts, while selecting quantities derived with 
reliable methods and good data. They also do not impose any interpretation 
on the results, such as K-corrections, evolutionary assumptions, or 
extinction estimates. Given the different instrumental and observational
limitations between the NIR and traditional optical bands, I have compiled 
(or applied corrections to obtain) the data for H-band and I-band. These are 
shown in the plots. 

   In addition to the studies referenced, a number of these are from 
ongoing and unpublished work of my own at the time
of writing. They include NIR and optical data from
CFHT with AO correction, and some HST WFPC2 data. The plots also show
for comparison, selected points for low redshift QSOs, which I think are
representative, but not at all complete, since there are hundreds of
low redshift QSO hosts in the literature. The plots also distinguish 
between radio-loud and radio-quiet objects, since this is likely to
be a fundamental and diagnostic difference. All points shown
are derived for H$_0$=50 and q$_0$=0 cosmology. This is for easiest comparison
rather than to represent a preferred choice.

\begin{figure}[ht]
\psfig{figure=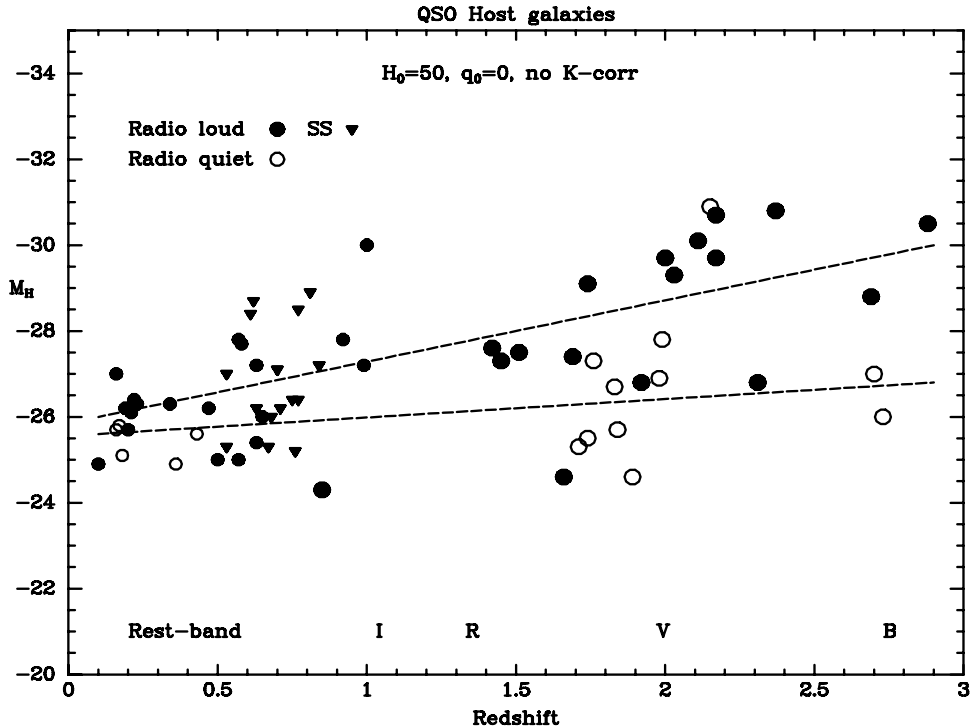,width=14cm,angle=90}
\caption{H-band absolute magnitudes of resolved host galaxies, as labelled.
All values are for given cosmology and have no K-correction, and some have
been derived from J or K band observations. z$<$0.3 points refer to
samples rather than individual objects, but are representative. The dotted
lines are linear fits to the radio-loud and radio-quiet data as plotted.
The rest wavelength bandpasses are given along the bottom.}
\end{figure}

\begin{figure}[ht]
\psfig{figure=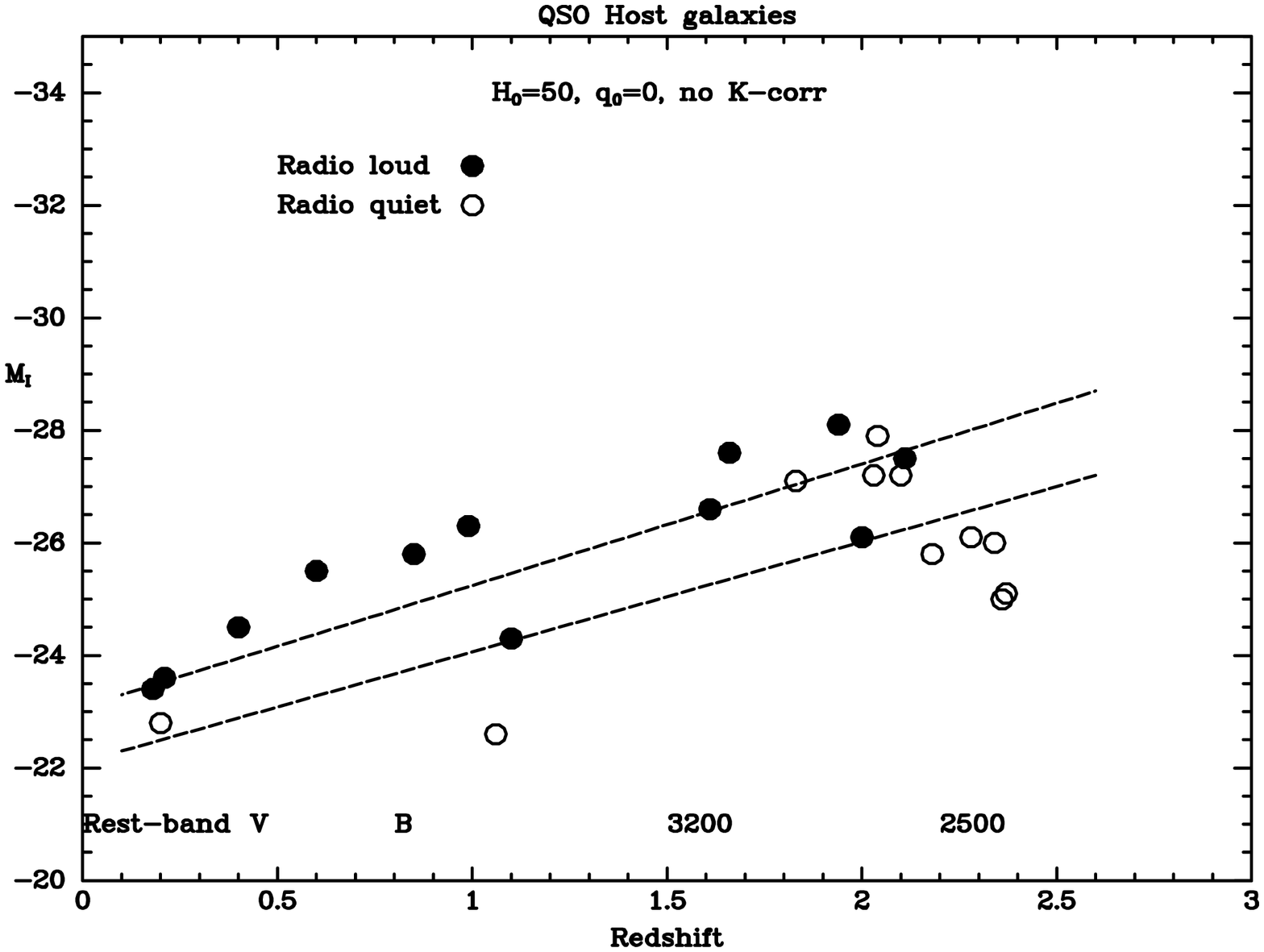,width=14cm,angle=90}
\caption{I-band absolute magnitudes as in figure 1}
\end{figure}

   While the comparison of
results for the same objects from different investigators are encouraging
in the few available instances, it is likely that both scatter and 
systematic differences arise from the different methods of removing PSFs, 
and overall image quality differences. I have omitted data that are less
reliable - usually from poor resolution, smaller telescopes, or more noisy
detectors. As a rough guide, the table shown compares the `quality'
of data from different sources. There are two important criteria
in data for host galaxy studies: detection of faint extended flux, and
image resolution to minimise the nuclear point source spread. Different
telescopes have different strengths in these two areas, and there are
additional criteria of detector noise and PSF complexity. Adding error bars 
to the diagrams to take all these into account would
be a large and unrewarding task, so I am assuming that the mix of 
different investigations in overlapping redshifts will show the overall
trends without severe systematic errors.

\begin{table}[ht]
\caption{Comparative power of investigations}
\begin{center}
\begin{tabular}{llrr|llrrr}

&~~~~NIR                                   &&&&~~~~CCD\cr
\hline
Telescope  &Res(")$^a$  &Det$^b$ &PSF$^c$ &Telescope &&Res(")  &Det &PSF\cr
\hline
 VLT     &0.5  &230  &50   &CFHT &HRCAM  &0.4  &12  &3\cr
 CFHT AO  &0.15    &91    &60    &&AO     &0.3  &47  &15\cr
 HST      &0.15   &100    &40   &HST  &WFPC2   &0.1  &4-20  &4-20\cr
 ESO      &0.8     &30     &4         &&STIS    &0.1    &30    &30\cr     
                                 &&&&KPNO       &&1.3   &17     &1\cr
                                 &&&&WHT        &&0.7   &88    &12\cr
\hline
\end{tabular}
\end{center}

$^a$ RES (") Image FWHM in arcsec

$^b$ Det = S/N achieved   (Aperture x thruput x exp / noise)

$^c$ PSF = Goodness of PSF removal  (Det x 0.1 / FWHM)

\end{table}

   Nothwithstanding these caveats, there are some clear points in the plots. 
Generally, the host galaxies are more luminous with increasing redshift. 
The increase is faster 
with redshift in visible wavelengths, and larger in the NIR in the 
radio-loud sources. However, the wavelength dependence is strongly
affected by the redshift itself, as well as evolution in the SED.
I have sketched in linear suggestions for the trends as observed. 
As noted by Falomo et al, at redshifts 1 and above, the radio-loud 
hosts are brighter than BCGs and L* galaxies, while radio-quiet ones
are comparable with BCGs. All are unusually bright galaxies, and there
a few enough null detections that this is unlikely to be an observational
issue.

    What can we infer from these results? Given that the observed rest
wavelength changes with redshift as shown along the X-axes, the host galaxies
are bluer as well as brighter, consistent with passive evolution or
declining star-formation as time proceeds, and the hosts at redshifts 2
and higher are starburst objects. This has been noted by various workers.
However, since QSO episodes are short compared with the lifetimes of
galaxies, we are not necessarily seeing the
evolution of a population of galaxies, but may also be looking at a changing
population of hosts with time. We come back to this below.

    As galaxy luminosity can reflect star-formation activity, increased
content by merging, or stripping by tidal events, we can remove some parameters
by looking at the average colours indicated by the lines through the 
H and I band plots. The colour plot
shows this evolution for the two QSO types. We show for comparison the
colour expected for constant star-formation, and passive evolution from
starburst for galaxy formed at redshift about 4, with the adopted
cosmology timescale. Dust and increased metallicity move the plots 
vertically up the diagram. 

\begin{figure}[ht]
\psfig{figure=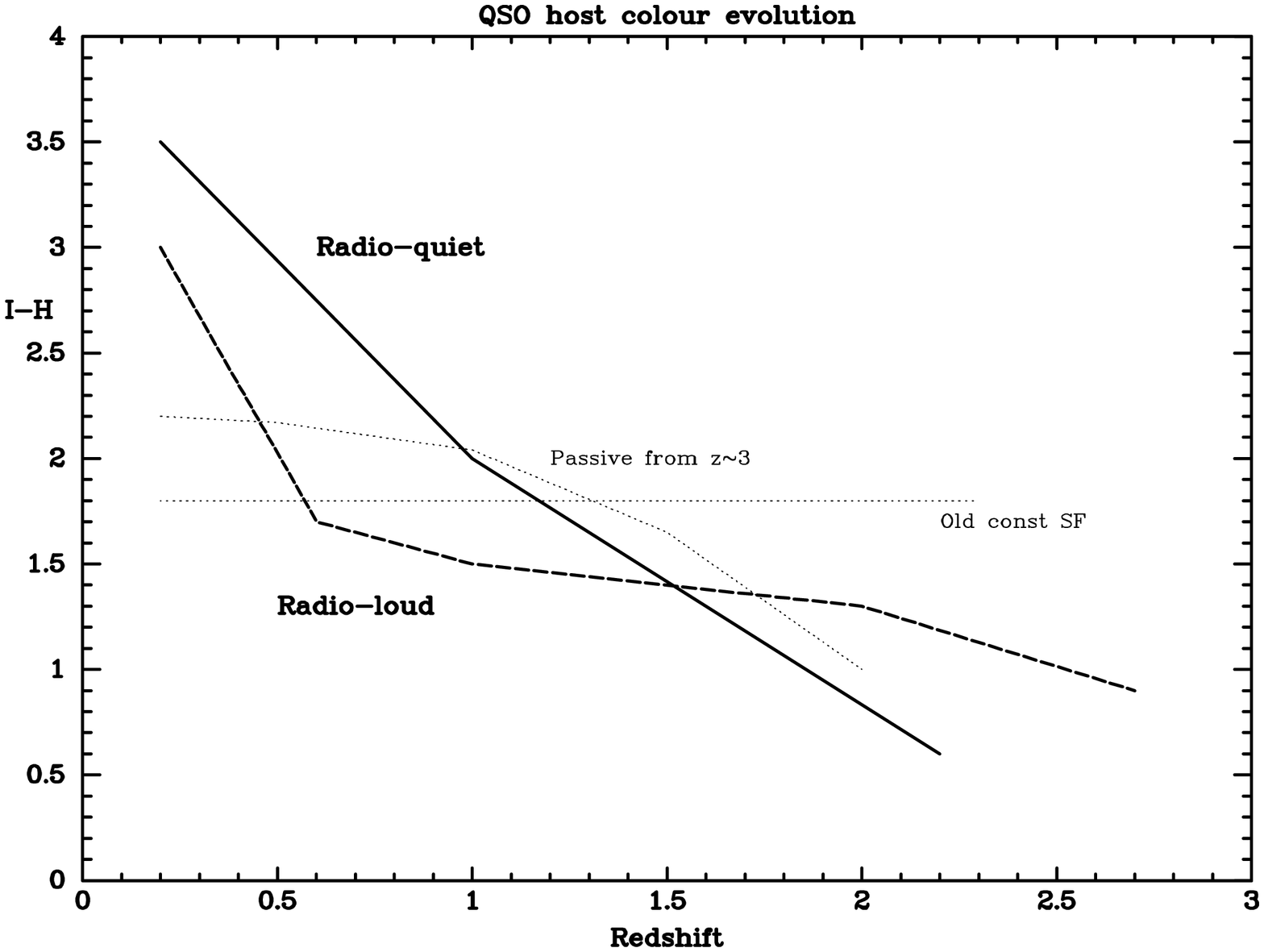,width=14cm,angle=90}
\caption{Locus of linear fits from Figures 1 and 2 converted to I-H.
The faint dotted lines are approximations from Gissel93 models for 
measured I-H as observed redshift changes. (The constant observed colour 
for constant star-formation arises because slow reddening is balanced by
the moving bandpasses.) Radio-loud hosts appear to arise in a population
that is actively forming stars over z=0.8 to 2, while radio-quiet hosts
have passivley evolving populations. The observed redder colours at low 
redshift may arise from dust or increasing metal abundance.}
\end{figure}

   The colour evolution is different for the two QSO types. At high
redshift, both have colours of star-forming galaxies, but the radio-loud ones
have an older population present. The radio-quiet hosts appear to evolve
passively to redshift 1, and then are reddened by abundance changes and/or 
dust. The radio-loud hosts appear to have active star-formation at high levels
down to redshift near 0.5, and then go the way of the radio-quiet.

\begin{figure}[ht]
\psfig{figure=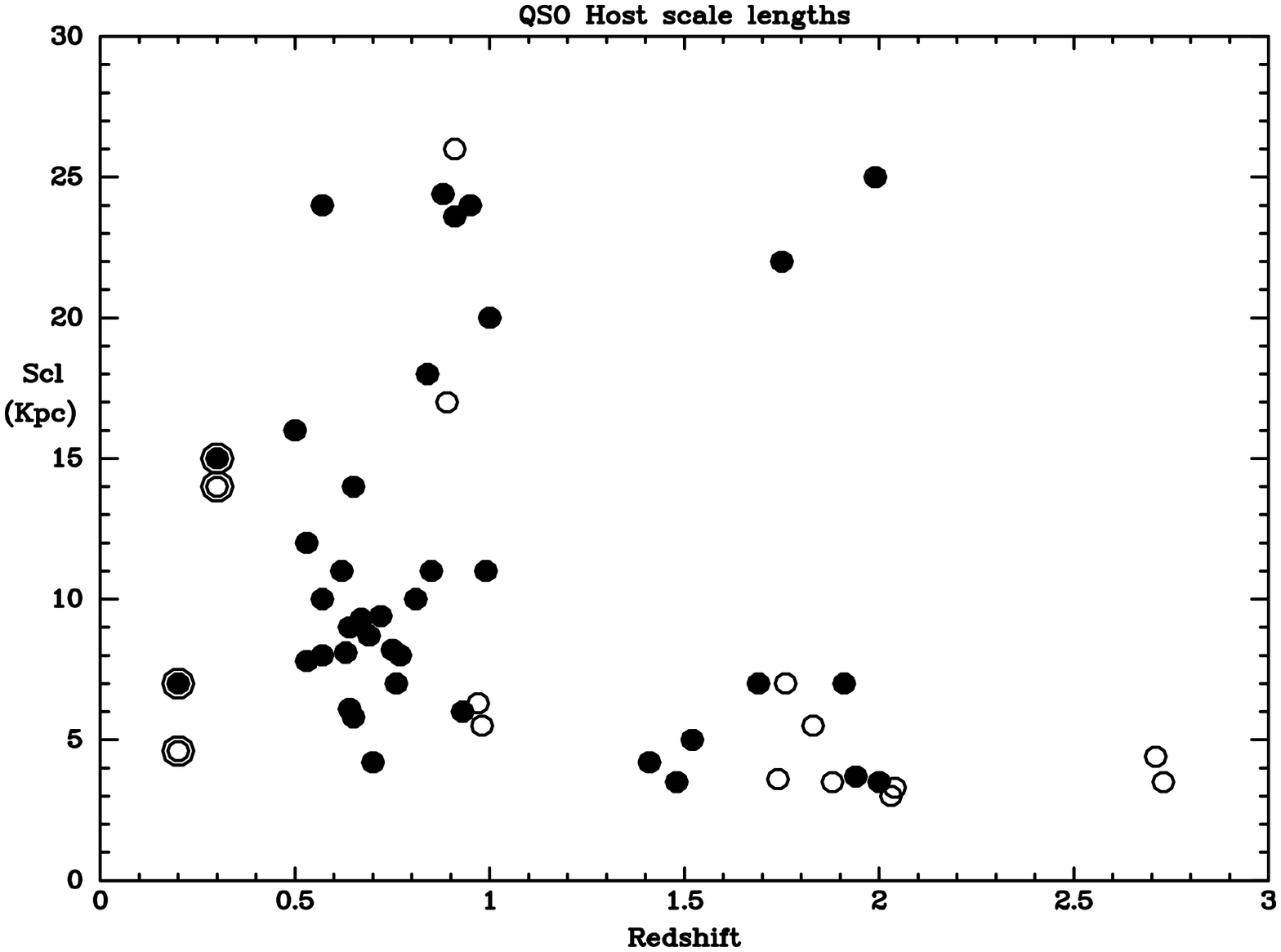,width=14cm,angle=90}
\caption{Host galaxy scale lengths for high redshift hosts, corrected for
differences in published values of FWHM, enclosed flux fraction. The two
low redshift points (circled) are for samples - the lower one being HST
short exposure images, which do not detect faint outer parts of the galaxies.}
\end{figure}

    The scale lengths are more subject to differences in PSF removal, but
the plot shows what the best data indicate. Generally, the high redshift
host galaxies are smaller, and become much smaller than present-day bright
galaxies. There is a scatter of high values that may arise from tidal tails.
If so, this is seen predominantly at redshifts 0.5 to 1. (when galaxy
merging is generally known to be higher). This is consistent with
heriarchical merging as galaxies evolve, as noted by other authors in the
subject. It is also consistent with the evolution of non-active galaxies
as seen by Lyman-break galaxy studies.

\section{Cluster environment}

   There have been several studies of QSO galaxy companions. Many authors
have noted the apparent connection between galaxy interactions and QSO
activity, as well the signs of past tidal events in the profiles of
QSO hosts at low redshift. These are much harder to see at high redshift, 
since tidal tails are old stars which are faint and close to the nuclear PSF.
It is simpler to count and even get spectra for nearby companions, in order
to characterise the galaxy environment of QSOs. However, such results are
quite dependent on image quality, signal level, and detection threshold. 
The table again shows where we expect to do best in this regard.

   The compendium of high redshift QSOs is shown in the Figure 5. The principal
references for these are Hutchings et al 1995, Hutchings 1995a, Teplitz et
al 1999, Wold et al 2000, 2001, Haines et al 2000, and some recent
unpublished data of my own. The numbers plotted
are estimated excess galaxies within 20 arcsec of the QSO in the sky,
in some cases derived from different areas originally reported.
The low-redshift values are means from many objects, and the vertical
`error bar' indicates the range among individual objects. A similar
range is revealed at redshift 1.1, which is QSOs in the `supercluster' 
region discussed by Hutchings et al (1995). The dotted lines sketch in suggested
increases in the RL and RQ environments above z=1. Clearly these are
very uncertain in view of the scatter where we have enough measures to know. 
However, it is generally accepted that QSOs do live in regions of enhanced
galaxy populations, and that the number of galaxy companions probably increases
at higher redshift. Several investigations show that the companion density
on the sky falls quickly with radius, so that the plot is oversimplified
in that respect.

\begin{figure}[ht]
\psfig{figure=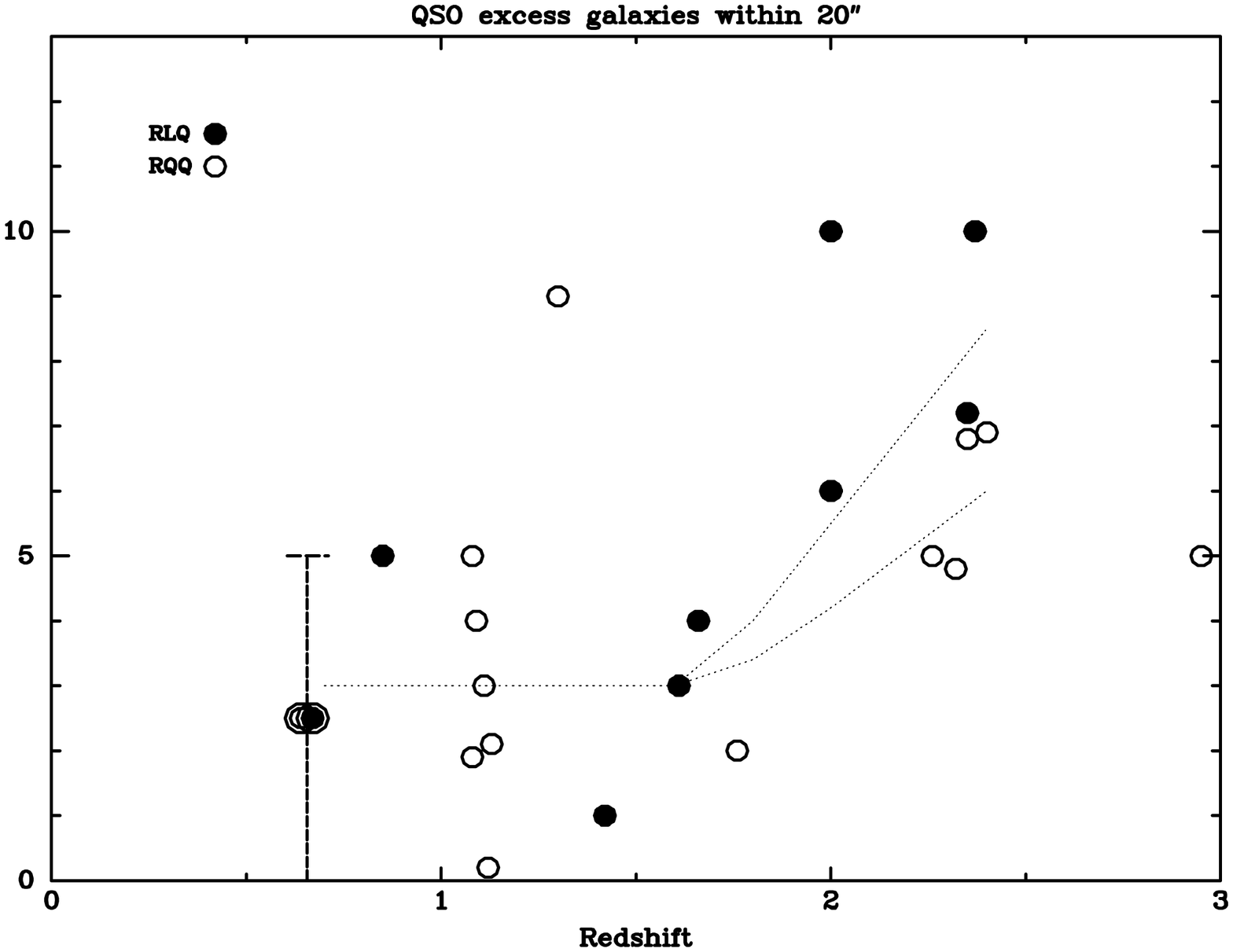,width=14cm,angle=90}
\caption{Counts of excess galaxies near QSOs. The values are corrected
to refer to the same area of sky, and for differences in limiting
magnitude. Dotted lines are suggested mean relationships for radio-loud and
radio-quiet QSOs.}
\end{figure}

   Another caveat is that at high redshift, the companions seen are compact,
so that their detection depends on good signal and high resolution. The 
deepest detection of companions comes from long CCD exposures with HST.
The small size of the faint galaxies means that HST can detect more of them,
to a magnitude fainter than CFHT, for example. However, at magnitude fainter
than 24,
most galaxies are high redshift star-forming objects, so that their association
with the QSO needs more than just blue colour from 2 filters. I have not 
plotted any of these very faint `companions' in the diagram.  

   In the NIR, HST has the advantage of dark sky, but 8m class telescopes 
have so much more light-gathering power that they win - not only for 
companions but also the faint parts of host galaxies that indicate tidal
events and evolved populations. However, no major investigations have been 
done yet.

    The galaxy companions we know about, are different at high redshift, and
suggestive of environments which may evolve into small groups (or a single 
large galaxy) at present time. The environment that triggers QSO activity 
almost certainly changes with cosmic time, ranging from initial merging of
protogalaxies to tidal triggers involving gas at present time. Radio-loud
sources are likely the more massive black hole (= galaxy bulge mass, 
age of galaxy, merging history,..?). It is also known in some cases that 
the companions are a foreground group, which may be connected with the QSO 
only by lensing action that makes the QSO more visible. We need more 
complete and careful investigations to get the required statistics on 
these issues.

\section{Line emission}

    The classic paper on line emission from AGN at high redshift is some
years ago, by McCarthy et al (1991). This established the `alignment effect'
between radio and emission-line structure that seems to be accepted
as the ionising effect of nuclear radiation and jet activity on
a gas-rich environment. For low redshift QSOs, there is another famous
paper by Stockton and MacKenty (1987) that investigates line emission, and
finds extended and irregular emission line gas predominantly in extended
radio-loud QSOs. This reveals the presence and complex dymanics of
gas that lies well outside the stellar population and is presumably
the result of tidal events that have fuelled the nucleus.

    Work on radio-quiet high redshift QSOs has occured only recently,
again with the help of larger telescopes, NIR detectors, and AO. Ohta
et al (2000) have detected [O II] emission in a z=4.7 QSO, and Hutchings et
al (2000) find [O III] and H$\alpha$ in a sample of radio-quiet or compact radio
source QSOs. These all have line emission that lies within the host galaxy
and thus corresponds to the NLR gas that has been mapped in low redshift 
Seyferts and QSOs. Where there is radio structure in these objects. the
line emission is aligned with it. The emission line images indicate
line equivalent widths typically 200 - 300\AA~ in the redshifted frame,
of which about half is typically from the nucleus.

   More detailed work will extend this work to a more statistically 
significant sample and perhaps lead to an understanding of the internal 
dynamics of the high redshift host galaxies.
 
\section{Summary}

   This review cannot do justice to the details of the different 
investigations. Other papers in this volume deal with individual
programs, which are included in my summary plots. 

   Overall, the work of the past decade has established a credible set of
investigations
of high redshift QSO hosts and environments. This should become more detailed
and extend to higher redshift with 8m class telescopes with AO, and advanced 
HST (and NGST) instrumentation. It seems well established that host galaxies
at high redshift are luminous, with active star-formation, and generally
are very compact for their luminosities. They also live in dense galaxy
companion environments. The hosts and their companions must undergo
significant merging and evolution, as the much less common present-epoch QSOs
are found in larger hosts with less crowded environments. We are on the
brink of studying the formation and evolution of the central black holes,
which can be measured by the host spheroid morphology, and the nuclear BLR
profiles. It is clear that the formation and triggering of QSOs is
an integral part of the formation of galaxies, and that QSOs at high redshift
offer a wealth of cosmological information for the years to come.






%


\bibliographystyle{apalike}

\clearpage
\centerline{References}

Aretxaga I., Terlevich, Boyle B.J., 1998, MNRAS, 296, 643

Falomo R., Kotilainen J.K., Treves A., 2001, ApJ, (astro-ph 0009181)

Hutchings J.B., 1995a, AJ, 109, 928

Hutchings J.B., 1995b, AJ, 110, 994

Hutchings J.B., 1998, AJ, 116, 20

Hutchings J.B., Crampton, D., Johnson A., 1995, AJ, 109, 73

Hutchings J.B., Crampton D., Morris S.L., Durand D., Steinbring E., 1999,
AJ, 117, 1109

Hutchings J.B., Morris S.L., and Crampton D., 2001, AJ, (astro-ph 0012245)

Haines C.P., Clowes R.G., Campusano L.E., 2001 (astro-ph 0012236)

Heckman T.M., Lehnert M.D., van Breugel W., Miley G.K., 1992, ApJ, 370, 78

Kotilainen J.K., Falomo R., Scarpa R., 1998, A\&A, 332, 503

Kotilainen J.K., Falomo R., 2000, A\&A, 2000, 364, 70

Kukula M.J. et al, 2001, MNRAS (astro-ph 0010007)

Lehnert M.D., Heckman T.M., Chambers K.C., Miley G.K., 1992, ApJ, 393, 68

Lehnert M.D., van Breugel W., Heckman T.M., Miley G.K., 1999, ApJS, 123, 351  

Lowenthal J.D., Heckman T.M., Lehnert M.D., Elias J.H., 1995, ApJ, 439, 588

McCarthy P.J., van Breugel W., Kapahi V.J., 1991, ApJ, 371, 478

Ohta K., et al, 2000, PASJ (astro-ph 0003107) 

Ridgway S.E., Heckman T.M., Calzetti D., Lehnert M., 2001, ApJ (astro-ph 0011330)

Rix H-W. et al, 2000, astro-ph 9910190

Stockton A., MacKenty J.W., 1987, ApJ, 316, 584

Teplitz H.I., McLean I.S., Malkan M.A., 1999, ApJ (astro-ph 9902231)

Wold M., Lacy M., Lilje P.B., Serjeant S., 2000, MNRAS, 316, 267

Wold M., Lacy M., Lilje P.B., Serjeant S., 2001, MNRAS (astro-ph 0011394)


\end{document}